\documentclass{appolb}

\usepackage{graphics}
\usepackage{graphicx}
\usepackage{epsfig}
\usepackage{color}		
\usepackage{amssymb}		

\usepackage{epsfig}
\usepackage{amsmath}
\usepackage{amsthm}
\usepackage{booktabs}
\usepackage{url}
\usepackage{longtable}
\usepackage{textcomp}
\usepackage{fontenc}
\usepackage{makeidx}
\usepackage{cases}
\usepackage{multicol}

\title{Impulsively Generated Linear and Non-linear Alfv\'en Waves in the Coronal Funnels}

\author{P. Chmielewski$^{a}$, A.K. Srivastava$^{b,c}$, K. Murawski$^{a}$, Z. Musielak$^{d,e}$\\
	\small{$^a$ Group of Astrophysics, UMCS, ul. Radziszewskiego 10, 20-031 Lublin, Poland.}\\
        \small{$^b$ Department of Physics, Indian Institute of Technology (BHU), Varanasi-221005, India.}\\
	\small{$^c$ Aryabhatta Research Institute of Observational Sciences (ARIES), Manora Peak, Nainital-263 129, India.}\\
	\small{$^d$ Department of Physics, University of Texas at Arlington, Arlington, TX 76019, USA.}\\
	\small{$^e$ Kiepenheuer-Institut f\"ur Sonnenphysik, Sch\"oneckstr. 6, 79104 Freiburg, Germany} \\
       }
\date{\today}

\begin{document}
\maketitle

\begin{abstract}
We present simulation results of the impulsively generated linear 
and non-linear Alfv\'en waves in the weakly curved coronal magnetic 
flux-tubes (coronal funnels) and discuss their implications for the 
coronal heating and solar wind acceleration.  We solve numerically 
the time-dependent magnetohydrodynamic (MHD) equations to obtain the temporal 
signatures of the small (linear) and large-amplitude (non-linear) Alfv\'en waves in the 
model atmosphere of expanding open magnetic field configuration (e.g., coronal funnels) 
by considering a realistic temperature distribution.  We compute the maximum 
transversal velocity of both linear and non-linear Alfv\'en waves
at different heights in the coronal funnel, and study their 
response in the solar corona during the time of their propagation.
We infer that the pulse-driven non-linear Alfv\'en waves may carry 
sufficient wave energy fluxes to heat the coronal funnels and also 
to power the solar wind that originates in these funnels.  Our study
of linear Alfv\'en waves show that they can contribute only to the 
plasma dynamics and heating of the funnel-like magnetic flux-tubes 
associated with the polar coronal holes.
\end{abstract}


\section{Introduction}

Small-amplitude (linear) Alfv\'en waves are incompressible magnetohydrodynamic 
(MHD) waves that propagate along the magnetic field lines, displacing them 
perpendicularly to the direction of wave propagation.  Alfv\'en waves carry  
energy from the solar sub-surface layers to its outer atmosphere [1]. 
Although, the theory of the linear Alfv\'en waves has been well established since 
the Nobel Prize discovery of Hannes Alfv\'en [2], 
the observations of 
these waves are difficult in the solar atmosphere due to their incompressible 
nature.  There is indirect evidence for the presence of these waves in the solar 
atmosphere given by the SOHO and TRACE observations.  However, more direct signature 
for the presence of such transverse waves (e.g., Alfv\'en waves) in different magnetic structures of the 
solar atmosphere at diverse spatio-temporal scales was supplied by the high-resolution observations recorded
by the Solar Optical Telescope (SOT) and the X-Ray Telescope (XRT) on board the Hinode Solar 
Observatory.  According to the discovery made by Okamoto et al. [3], De Pontieu et al. [4] and Cirtain et al. [5],
the signature 
of Alfv\'en waves were observed in prominences, spicules and X-ray jets respectively
in the solar atmosphere using these instruments.  Interpretations of these observations in terms of 
Alfv\'en waves is still not commonly accepted, see for example 
Erd{\'e}lyi \& Fedun [6], Van Doorsselaere et al. [7] and Goossens et al. [8,9]
who concluded that the observed waves were rather 
magnetoacoustic kink waves instead of being pure incompressible
Alfv\'en waves.  However, more recent arguments by Goossens et al. [9]
seem to clearly indicate that the original interpretation of the observations in terms
of incompressible Alfv\'en waves was indeed correct.  Additional observational 
support was given by Tian et al. [10] 
who spectroscopically detected Alfv\'en waves.

Observational signature for the existence of purely incompressible torsional Alfv\'en waves in the 
confined magnetic fluxtube of the lower solar 
atmosphere was also reported by Jess et al. [11],
who analyzed the H$\alpha$ observations 
recorded with high spatial resolution by the Swedish Solar Telescope (SST).  They 
interpreted and described the unique observations SST in terms of Alfv\'en waves in the localized chromosphere with 
periods from 12 min down to the sampling limit of the recorded observations near 2 min, 
with maximum power near 6-7 min. Tomczyk et al. [12] 
also reported on the ubiquitous 
presence of the Alfv\'en waves in the large-scale corona using the ground based 
observations of Coronal Multi-channel Polarimeter.

Extensive theoretical and observational studies of Alfv\'en waves are important
in the context of te Sun because these waves are likely candidates for the atmospheric heating as well as
supersonic wind acceleration in its polar coronal holes [13-15]. 
In such open field regions on the polar cap of the Sun, the non-thermal broadening 
of spectral lines was observed [16-21], 
which indicates the process of exchanging energy from Alfv\'en waves to the ambient plasma 
does occur in these regions of the solar atmosphere.  Recently, Chmielewski et al. [22] 
modelled numerically the impulsively excited non-linear Alfv\'en waves in solar coronal holes, 
and concluded that such waves may result in the observed spectral line broadening in 
the solar atmosphere [17]. 
Zaqarashvili et al. [23] 
have also reported that the resonant 
energy conversion from Alfv\'en (transverse) to acoustic (longitudinal) waves in the lower solar atmosphere where plasma $\beta$ 
becomes unity. This effect can be responsible for the spectral line 
width variation and narrowing due to the dissipation of Alfv\'en waves.  However, this theory could only demonstrate the most likely 
physical scenario behind the line-width reduction as observed only by O'Shea et al. [19] 
in polar coronal hole 
beyond 1.21 solar radii. Additionally, there is not enough 
observational evidences that provide the resonant energy conversion in the solar corona
between Alfv\'en as well as magnetoacoustic waves [24-25].

The above described recent discoveries of Alfv\'en waves in the solar atmosphere 
well-justify extensive studies of the waves that were performed by numerous 
investigators in the last few decades both analytically and numerically [26-28]. 
These studies covered both linear [29] 
and non-linear [30-32] 
Alfv\'en waves, and different aspects of the wave 
generation, propagation and dissipation were investigated.  The specific objectives 
of these studies were to understand the role of Alfv\'en waves in the atmospheric 
heating and localized plasma dynamics.  In this paper, we numerically investigate 
the behavior of small-amplitude (linear) and large-amplitude (non-linear) Alfv\'en 
waves in a model of the solar atmosphere with the weakly curved magnetic field that 
mimics open magnetic field structure of the solar corona.  Such magnetic field 
configurations, which are commonly known as coronal funnels, can be widely applicable
to the polar coronal holes, expanding open arches near the boundary of the coronal 
holes in the quiet-Sun as well as fan-loop arches, where plasma and plasma flows 
are observed.  In Sec.~2, we discuss the numerical simulation of pulse-driven 
Alfv\'en waves.  We present the results of numerical simulations in Sec.~3.  
Our discussion of the obtained results and conclusions are given in the last 
section of this paper.

\section{Numerical Simulation of the Pulse-driven Alfv\'en Waves in Coronal Funnels}

Our model of the solar atmosphere consists of a gravitationally-stratified 
plasma in a weakly curved magnetic field configuration that mimics the 
coronal funnels in the solar atmosphere. The model is governed by 
the following set of ideal magnetohydrodynamic (MHD) equations:

\begin{equation}
\label{eq:MHD_rho} 
{{\partial \varrho}\over {\partial t}}+\nabla \cdot (\varrho{\bf V})=0\, ,\\
\end{equation}

\begin{equation}
\label{eq:MHD_V}
\varrho{{\partial {\bf V}}\over {\partial t}}+ \varrho\left ({\bf V}\cdot \nabla\right ){\bf V}= 
-\nabla p+ \frac{1}{\mu} (\nabla\times{\bf B})\times{\bf B} +\varrho{\bf g}\, , \\
\end{equation}

\begin{equation}
\label{eq:MHD_B}
{{\partial {\bf B}}\over {\partial t}}= \nabla \times ({\bf V}\times {\bf B})\, , \\
\end{equation}

\begin{equation}
\label{eq:MHD_divB}
\nabla\cdot{\bf B} = 0\, , \\
\end{equation}

\begin{equation}
\label{eq:MHD_p}
{\partial p\over \partial t} + {\bf V}\cdot\nabla p = -\gamma p \nabla \cdot {\bf V}\, ,\\
\end{equation}

\begin{equation}
\label{eq:MHD_CLAP}
p = \frac{k_{\rm B}}{m} \varrho T\,
\end{equation}
Here ${\varrho}$ is the mass density, ${\bf V}$ and ${\bf B}$ are respectively the vectors of
the flow velocity and the magnetic field.
Moreover, $p$, $\gamma=5/3$, $\mu$, ${\bf g}=(0,-g,0)$, $T$, $m$, $k_{\rm B}$, are respectively
the gas pressure, adiabatic index, 
 the magnetic permeability, a vector of gravitational acceleration with 
its value $g=274$ m s$^{-2}$, temperature, a mean particle mass, and a Boltzmann's constant.

We consider a 2.5D model of the solar atmosphere with an invariant coordinate ($\partial/\partial z = 0$)
and varying the $z$-components of velocity ($V_{\rm z}$) and magnetic field ($B_{\rm z}$)
with $x$ and $y$.
The solar atmosphere is in static equilibrium 
(${\bf V}_{\rm e}={\bf 0}$) with force- and current-free magnetic field, i.e.,
\begin{equation}
(\nabla\times{\bf B}_{\rm e})\times{\bf B}_{\rm e} = {\bf 0}\, , \hspace{4mm} 
\nabla\times {\bf B}_{\rm e}={\bf 0}\, .
\label{eq:B}
\end{equation}
Equilibrium quantities are described by the subscript $_{\rm e}$.

In our model of atmosphere a curved magnetic field of the coronal funnel
\begin{equation}
\label{eq:B_e}
{\bf B}_e = \nabla\times{\bf A_{\rm e}}\, ,
\end{equation}
is given by the following magnetic flux function{\bf :}
\begin{equation}
\label{eq:A}
{\bf A_{\rm e}} = \Lambda_{\rm B} B_{\rm 0} \cos(\frac{x} {\Lambda_{\rm B}})\exp{(-\frac{y-y_{\rm r}} {\Lambda_{\rm B}})} \, {\bf\hat{z}}\, ,
\end{equation}
where ${\bf\hat{z}}$ is a unit vector along the $z$-direction and
$B_{\rm 0}$ is the magnetic field at the reference level, $y=y_{\rm r}$, 
which is chosen at $y_{\rm r}=10$ Mm in our numerical simulation setup.  We set and hold fixed $B_{\rm 0}$ 
in such a way that the Alfv\'en speed, $c_{\rm A}=B_{\rm 0}/\sqrt{\mu 
\varrho_{\rm e}(y=y_{\rm r})}$ is ten times greater than the sound speed, 
$c_{\rm s}=\sqrt{\gamma p_{\rm e}(y=y_{\rm r})/\varrho_{\rm e}(y=y_{\rm r})}$
in the model atmosphere. 
Such a choice of ${\bf B}_{\rm e}$ results in Eq.~(\ref{eq:B})is being satisfied
during the numerical simulation.
Here $\Lambda_B=2L/\pi$ denotes the magnetic scale-height and $L$ is a half 
of the magnetic arcade width as considered in our model. Since we aim to model a weakly expanding coronal 
funnels, we take $L=75$ Mm and keep it fixed in our numerical model.  For this 
setting, the magnetic field lines are weakly curved and represent the open 
and expanding field lines similar to the coronal funnels and arches in the 
real solar atmosphere.

As a result of the implementation of Eq.~(\ref{eq:B}), the pressure gradient is balanced by the 
gravitational force in the model atmosphere, i.e., 
\begin{equation}\label{eq:p}
-\nabla p_{\rm e} + \varrho_{\rm e} {\bf g} = {\bf 0}\, .
\end{equation}
Using the ideal gas law of Eq. (\ref{eq:MHD_CLAP}) and the $y$-component of 
the hydrostatic pressure balance given by Eq. (\ref{eq:p}), we express
the equilibrium gas pressure and mass density in the set model atmosphere for the coronal funnels as
\begin{equation}\label{eq:pres} 
p_{\rm e}(y)=p_{\rm 0}~{\rm exp}\left( -
\int_{y_{\rm r}}^{y}\frac{dy^{'}}{\Lambda (y^{'})} \right),\hspace{4mm}
\varrho_{\rm e} (y)=\frac{p_{\rm e}(y)}{g \Lambda (y)}\, ,
\end{equation}
where
\begin{equation} 
\Lambda(y) = \frac{k_{\rm B} T_{\rm e}(y)} {mg}
\end{equation}
is the pressure scale-height, and $p_{\rm 0}$ denotes the gas pressure at the chosen
reference level. 

\begin{figure}
\begin{center}
\mbox{
\includegraphics[width=6.5cm]{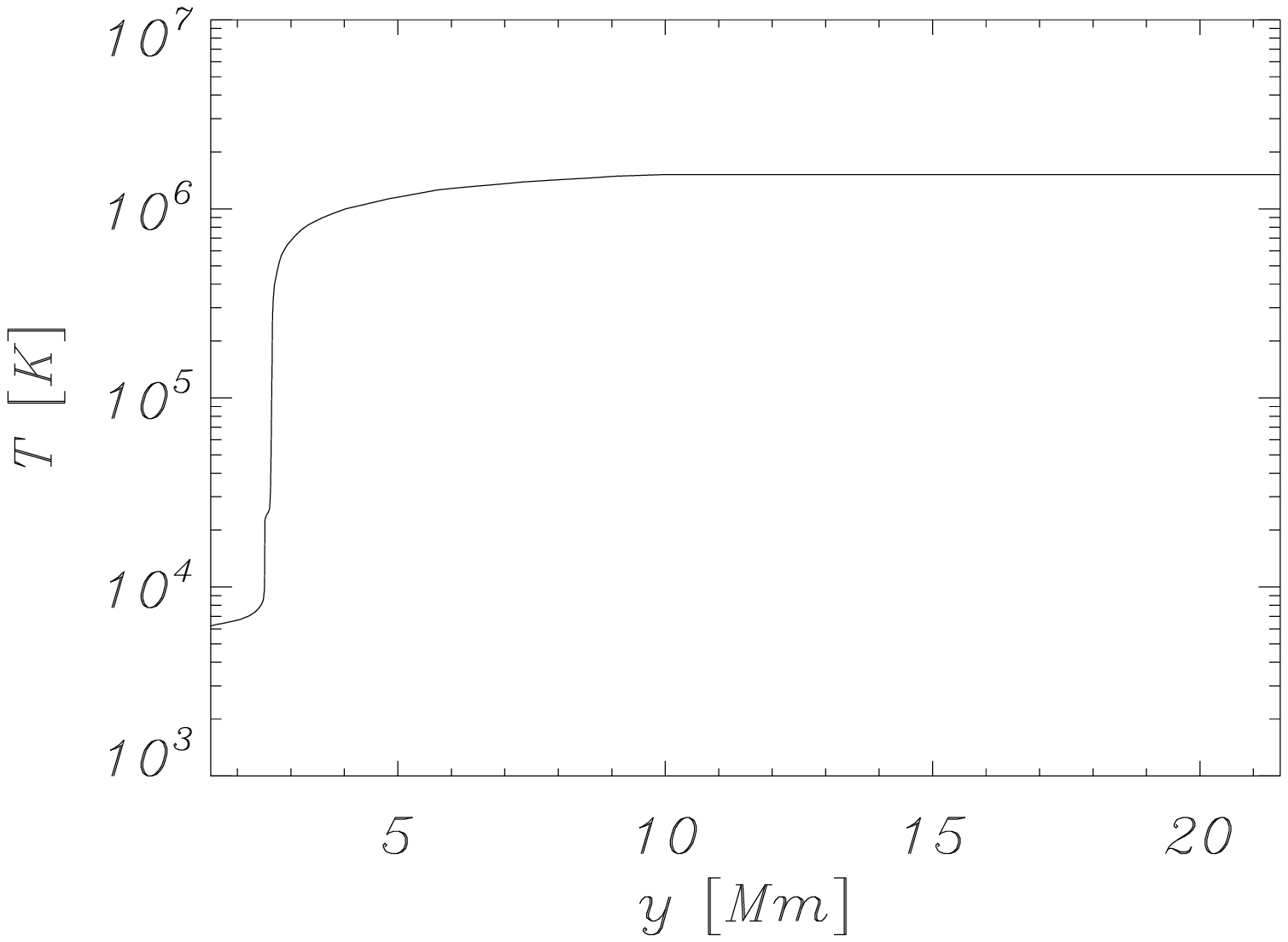}
\includegraphics[width=6.5cm]{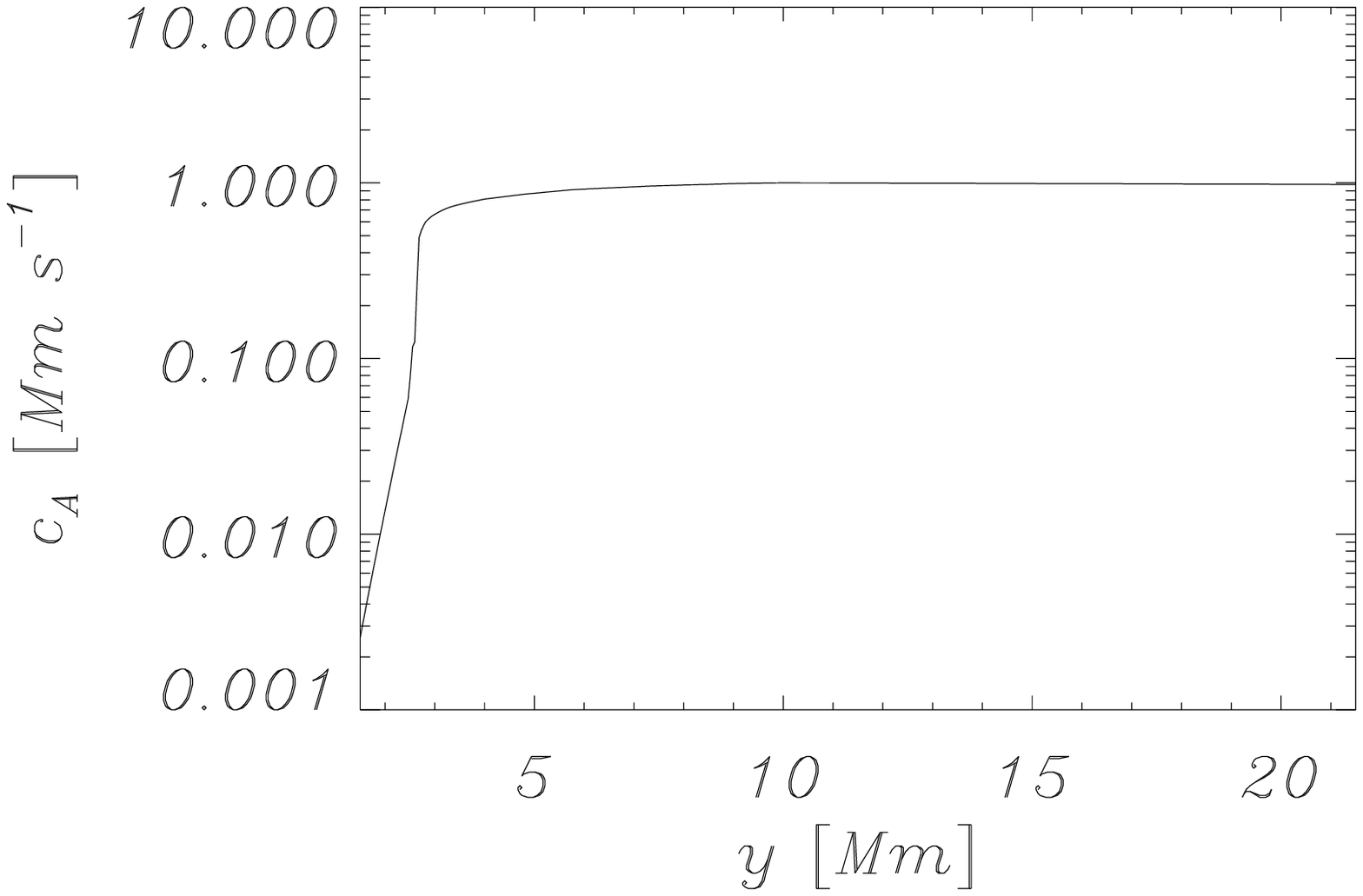}}
\caption{\small 
Equilibrium profile of the temperature (left panel) and the Alfv\'en speed (right panel)
in the model solar atmosphere of the coronal funnels.
} 
\label{fig:Tealf}
\end{center}
\end{figure}

We consider a realistic model of the plasma temperature profile in the model 
atmosphere of the coronal funnels [33] 
as displayed in Fig.~\ref{fig:Tealf} (left panel).  Temperature attains a value of about 
$6 \times 10^{3}$~K at $y=1.5$ Mm and it increases up to $1.5 \times 10^{6}$ K 
in the solar corona at $y=10$ Mm in the higher parts of the model
atmosphere.  In the higher solar corona, the temperature is assumed to be constant typically at
million degree Kelvin.
The temperature profile determines the equilibrium mass density and gas pressure 
profiles in the model solar atmosphere.  Both $\varrho_{\rm e}(y)$ and $p_{\rm e}(y)$ experience a sudden drop
in their typical values at the transition region that is located at $y \simeq 2.7$ Mm
in our numerical setup.
%
%

In this model of the coronal funnels, the Alfv\'en speed, $c_{\rm A}$ varies only with $y$ and is expressed as follows: 
\begin{equation}
\label{eq:ca}
c_{\rm A}(y) = \frac{B_{\rm 0}e^{-\frac{y-y_{\rm r}}{\Lambda_{\rm B}}}}{\sqrt{\mu \varrho_{\rm e}(y)}}\, .
\end{equation}
The Alfv\'en velocity profile  
is displayed in Fig.~\ref{fig:Tealf} (right panel), while it should be noted that the Alfv\'en speed in the chromosphere, 
$c_{\rm A}(y=1.75\, {\rm Mm})$, is about $25$ km s$^{-1}$.
The Alfv\'en speed rises abruptly through the transition region reaching a value 
of $c_{\rm A}(y=10\, {\rm Mm}) = 10^{3}$ km s$^{-1}$ (Fig.~\ref{fig:Tealf}, right panel) in the 
model atmosphere.
The increment of $c_{\rm A}(y)$ with height results from a faster decrement of $\varrho_{\rm e}(y)$
than $B_{\rm e}(y)$ with the height in the coronal funnel. It is worth to mention that 
this profile is more realistic than the one used by Murawski \& Musielak [29].

The realistic solar atmosphere above the polar coronal holes, the fan-like arches 
near the boundary of active regions, quiet-Sun expanding flux-tubes, reveal  
the complexity of its plasma and magnetic field structure.  The magnetic field 
configuration in such weakly expanding magnetized structures can be approximated  
by expanding coronal funnels in the lower part of their atmosphere and comparatively 
smooth and homogeneous open field lines in their upper parts [34,~35]. 
This indicates that
the field structure and magnetic scale-height vary in such coronal funnels on wider aspects, and 
the field configuration changes from dipolar to multipolar during the changes 
from the solar minimum to its maximum. In the above viewpoints and inspite of these well-known variations, 
we although assume that the magnetic field scale height is fixed at a reasonable value 
of 75 Mm in our numerical model that represents the weakly curved open 
field lines of the coronal funnels in the solar atmosphere. 
This assumption does not affect the validity of our 
numerical results as these weakly expanding coronal funnels can extend up to 
few hundreds megameters, while they can be few megameter wide in the horizontal 
direction [34,~35].

Equations considered in our simulation model (\ref{eq:MHD_rho})-(\ref{eq:MHD_CLAP}) are solved
numerically with a use of the FLASH code [36], which 
implements a second-order unsplit Godunov solver with various slope
limiters and Riemann solvers 
as well as Adaptive Mesh Refinement
(AMR). 
We use the minmod slope limiter and the
Roe Riemann solver [37] as implemented in the FLASH code. 
The two dimensional simulation box is set as
$(-5\, {\rm Mm},5\, {\rm Mm}) \times (1\, {\rm Mm},81\, {\rm Mm})$ 
and there is imposed fixed in time boundary conditions for all plasma quantities
in $x$- and $y$-directions. All plasma quantities remain invariant along the $z$-direction.
However, $V_{\rm z}$ and $B_{\rm z}$, in general, are different than zero.
In the present numerical simulation, we use AMR grid with a minimum (maximum) level of
refinement set to $3$ ($8$), while this refinement strategy was based on
controlling the numerical errors in mass density during the numerical modelling. 

Now, in order to get a fine grid, our simulation region is enshroud
by eight equilateral blocks that corresponds to zeroth level of grid refinement.
Then, some these blocks are divided into $2^n$ smaller blocks,
where $n=2$ denotes space dimension of our numerical model.
In this way, we raise the level of refinement
and we get an initial, non-uniform numerical grid. 

Our simulation box is covered by the numerical grid,
which is finer below the transition region
and along $x=0$, the path of Alfv\'en wave propagation, 
and rare, but dense enough in the solar corona.

%
This results in an excellent resolution of steep spatial profiles and
greatly reduces the numerical diffusion at
these locations in the simulation domain.
%
As every numerical block consists of $8\times 8$ identical numerical cells,
we reach the finest spatial resolution of 39~km.
Note that our simulation region is chosen large and
enough to observe wide pulse Alfv\'en waves propagating along magnetic 
field lines.
Therefore, we aim to model a general coronal funnel structures 
where such waves are evolved.

We perturb initially (at $t=0$ s) the equilibrium model atmosphere mimicing the coronal funnels
by a Gaussian pulse in the $z$-component of velocity given by 
\begin{equation}
\label{eq:init_per}
V_{\rm z}(x,y,t=0) = A_{\rm v} \exp\left[ -\frac{(x-x_{\rm 0})^2+(y-y_{\rm 0})^2}{w^2} 
\right]\, ,
\end{equation}
where $A_{\rm v}$, $(x_{\rm 0},y_{\rm 0})$, and $w$, are respectively the amplitude of the pulse, its 
initial position, and width. 
We set $w=0.2$ ${\rm Mm}$, 
$(x_{\rm 0}=0, y_{\rm 0}=1.75)$ Mm and consider 
three cases: (a) $A_{\rm v}=5$ km~s$^{-1}$; (b) $A_{\rm v}=10$ km~s$^{-1}$;
{(c) $A_{\rm v}=40$ km~s$^{-1}$} for our numeircal calculations.
It should be noted that in  the 2.5D model developed by us, the Alfv\'en waves decouple from the magnetoacoustic waves
and it can be described by $V_{\rm z}(x,y,t)$.
In the consequence of it, the initial pulse triggers Alfv\'en waves which in their linear limit can be described by the wave equation
\begin{equation}
\label{eq:Vz_lin}
\frac{\partial^2 V_{\rm z}}{\partial t^2} = c_{\rm A}^2(y) \frac{\partial^2 V_{\rm z}}{\partial y^2}\, .\\
\end{equation}
%

\section{Results of the Numerical Simulation of Alfv\'en Waves in Coronal Funnels}

We simulate both linear ($A_{\rm v}=5$~km s$^{-1}$) and non-linear ($A_{\rm v}=10$~km s$^{-1}$ 
and {$A_{\rm v}=40$~km s$^{-1}$}) impulsively excited Alfv\'en waves and study their 
propagation along the open and expanding magnetic field lines of the coronal funnels. 
We categorize linear (non-linear) waves as those whose amplitude is significantly 
smaller, by 10\%, than (or comparable to) a local Alfv\'en velocity.  The amplitude 
with few {kilometers} per second will generate the linear Alfv\'enic perturbations.  
While the amplitude with few tens of kilometer per second is responsible for the 
non-linear Alfv\'en waves.
%
The results are compared to those 
previously obtained by Chmielewski et al. [22].  It should be noted that the effect of 
inhomogeneities in the magnetic field as well as plasma properties across the magnetic field lines 
of the coronal funnel are not included in our study.  The 
Alfv\'en waves are generated by the transversal velocity pulse perpendicular to the magnetic 
isosurface (X-Y) in the $z$-direction and we compute the maximum transversal velocity 
($V_{\rm z}$) at different heights in the simulation domain. 

\begin{figure}
\centering
{
\includegraphics[width=8.25cm,angle=0]{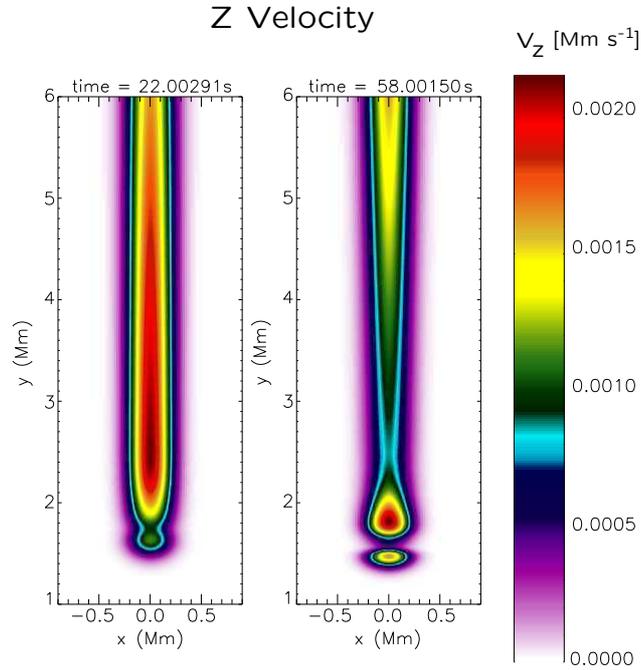}
}
\caption{\small
Transverse velocity $V_{\rm z}$ profiles 
at {$t=22$ s} 
and $t=58$ s for 
$A_{\rm v}=5$ km s$^{-1}$. 
}
\label{fig:6panels}
\end{figure}

\begin{figure*}
\begin{center}
\mbox{
\hspace{-2.0cm}
\includegraphics[width=5.5cm]{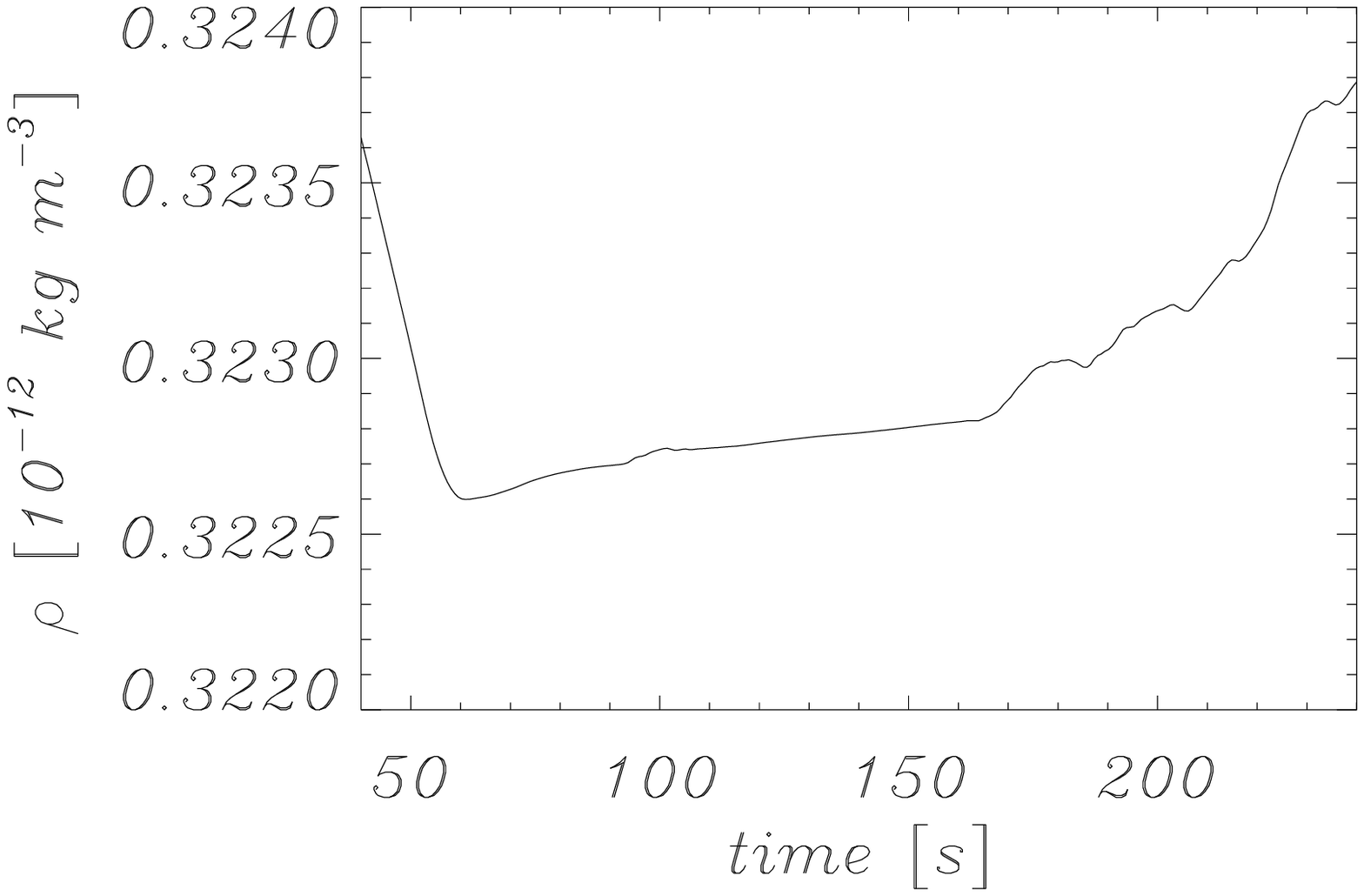}
\includegraphics[width=5.5cm]{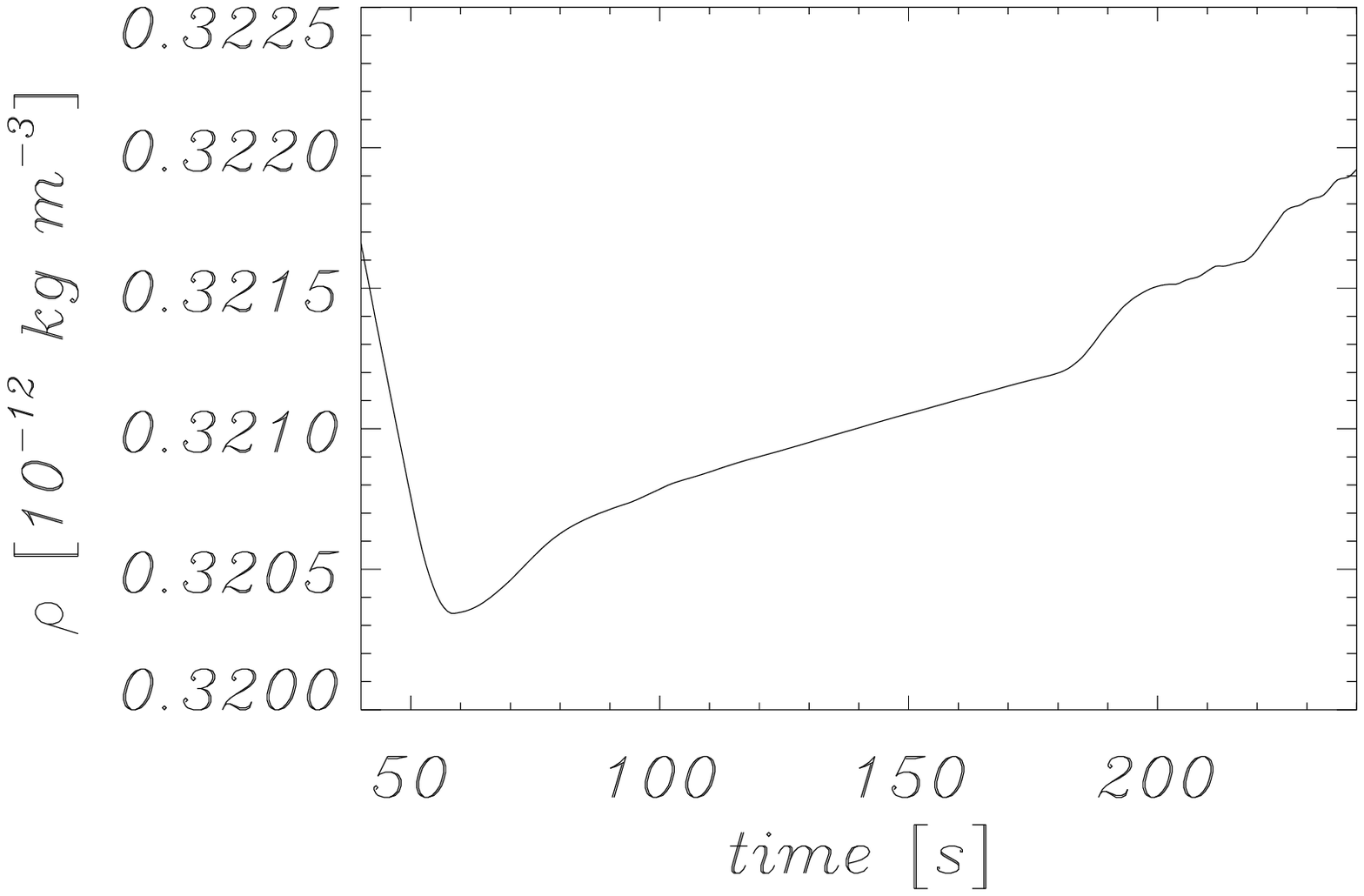}
\includegraphics[width=5.5cm]{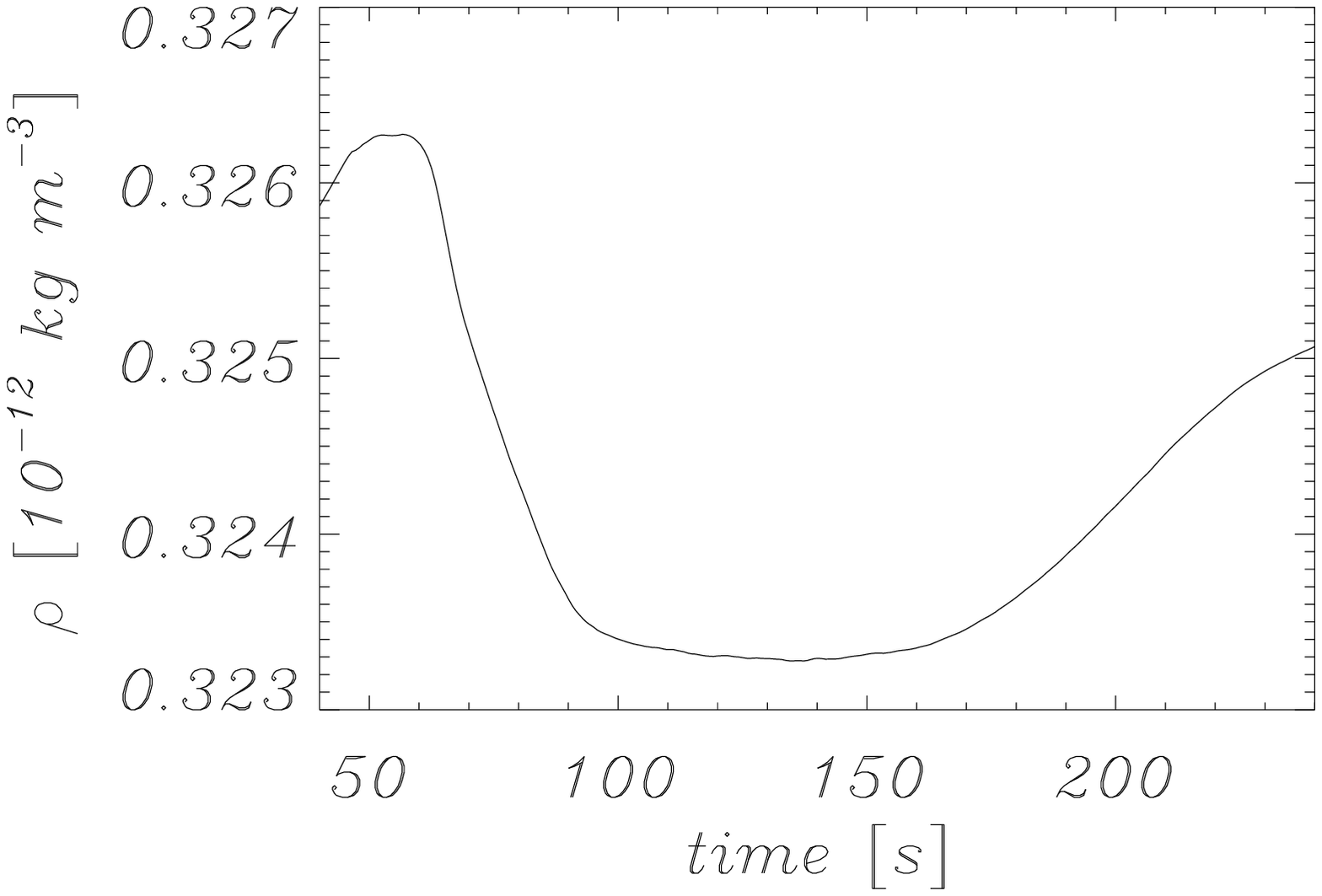}
}
\mbox{
\hspace{-2.0cm}
\includegraphics[width=5.5cm]{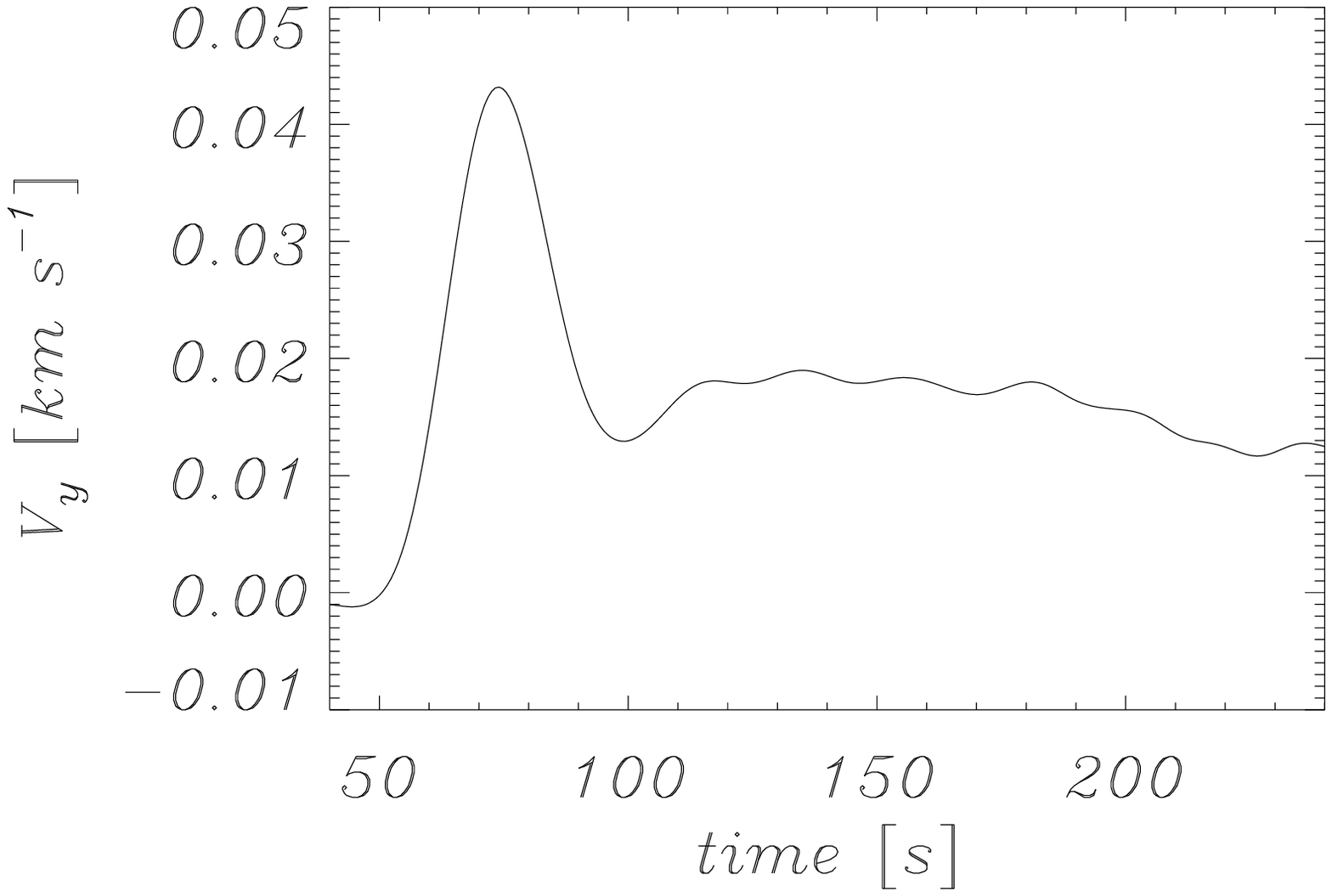}
\includegraphics[width=5.5cm]{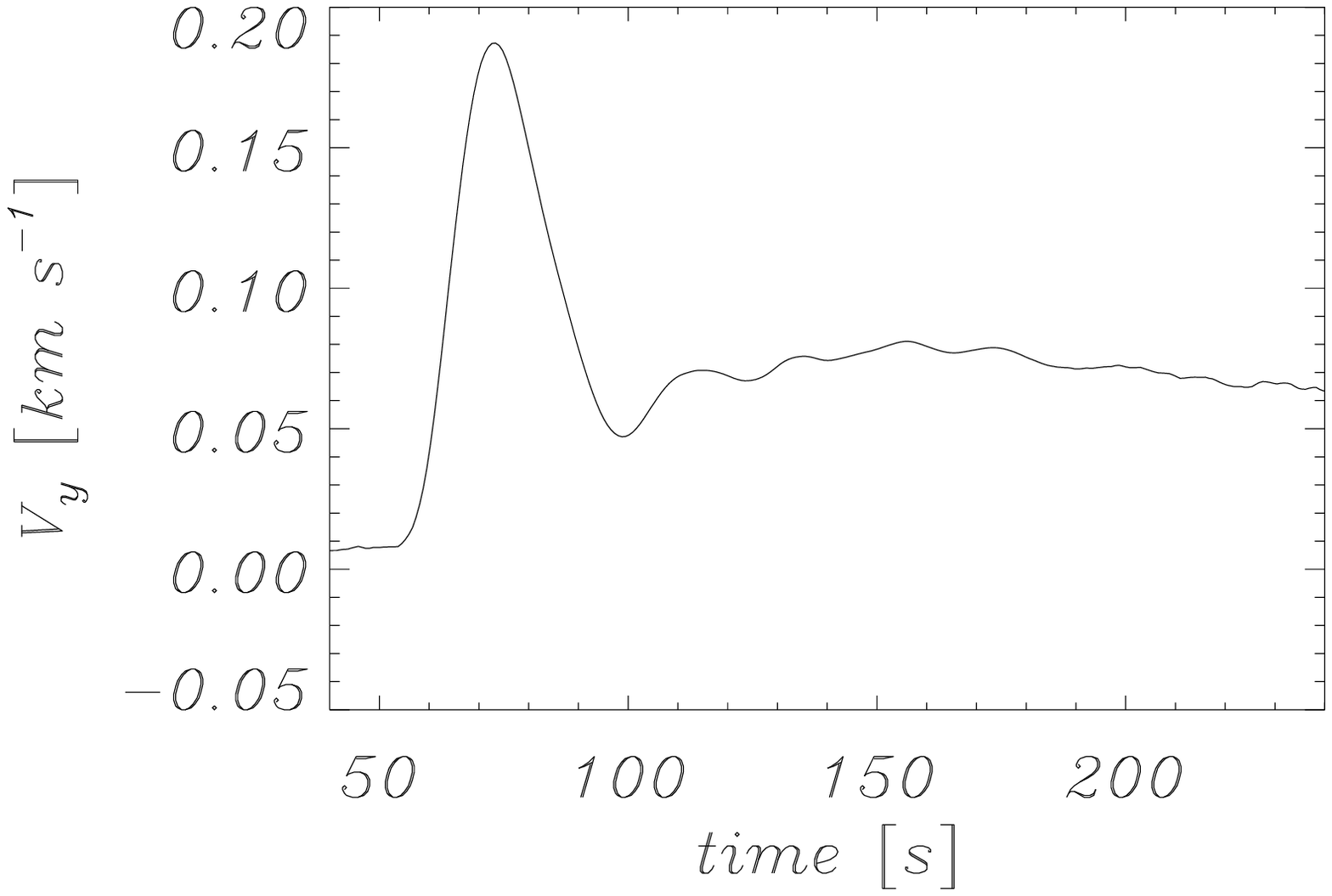}
\includegraphics[width=5.5cm]{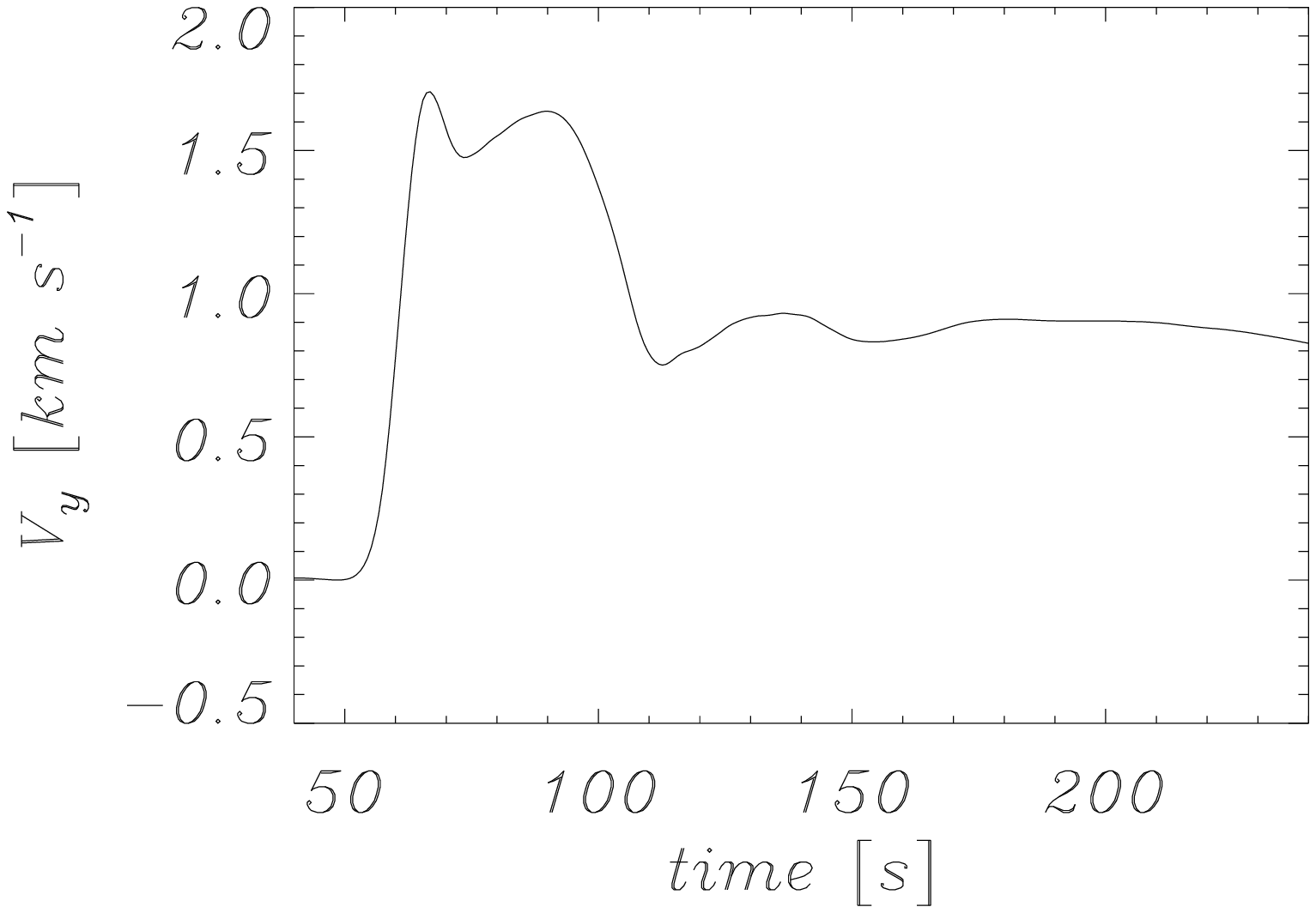}
}
\mbox{
\hspace{-2.0cm}
\includegraphics[width=5.5cm]{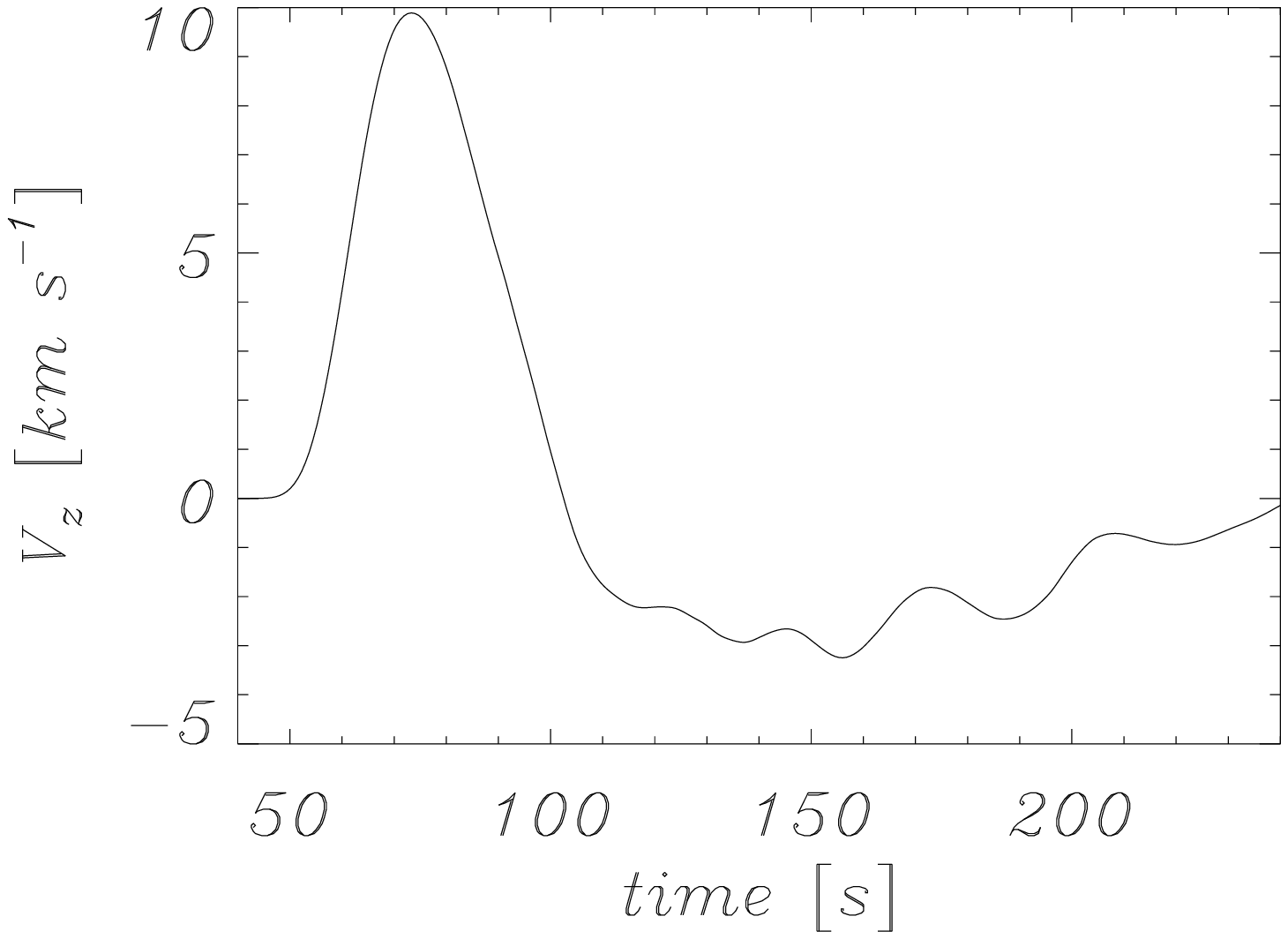}
\includegraphics[width=5.5cm]{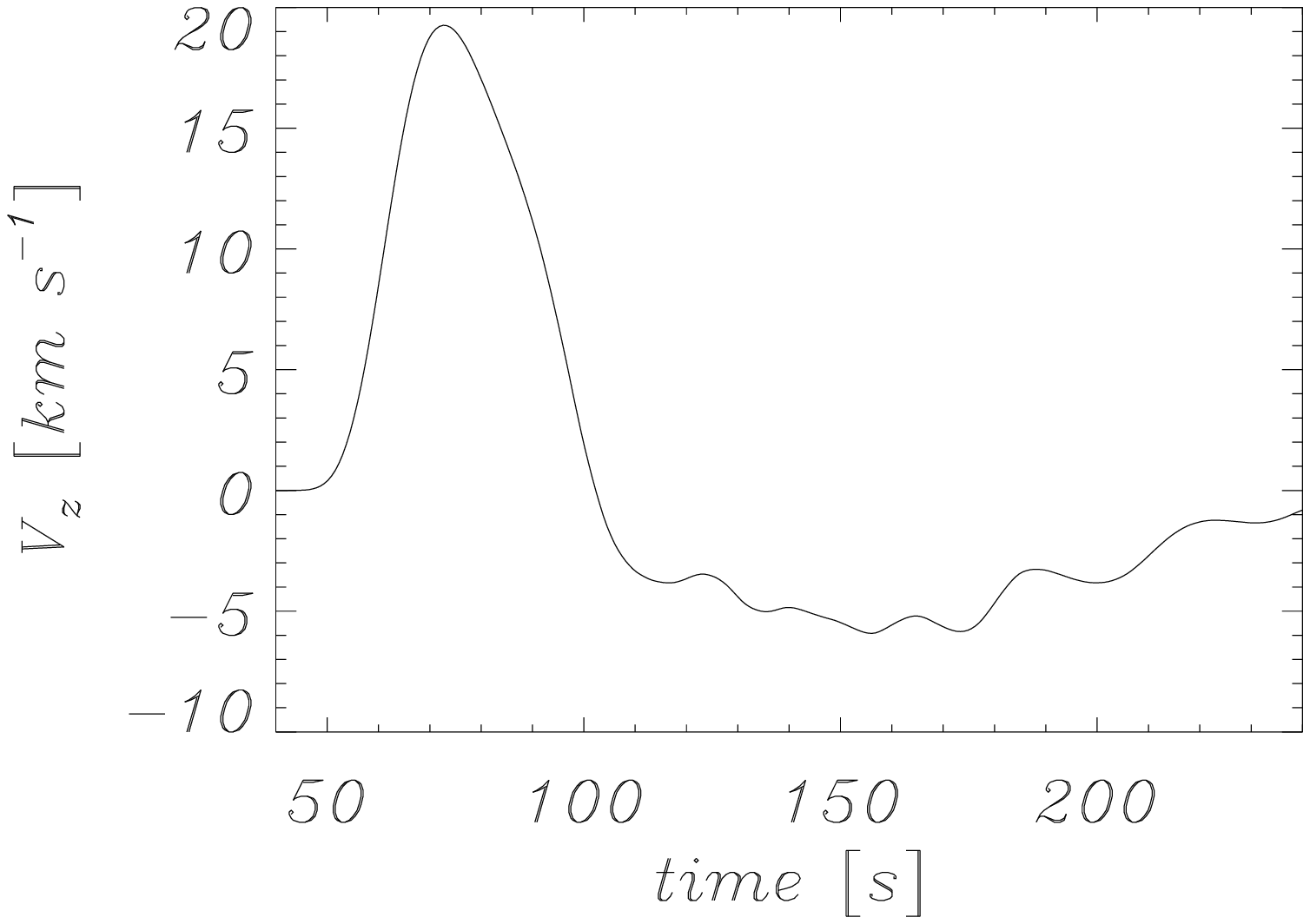}
\includegraphics[width=5.5cm]{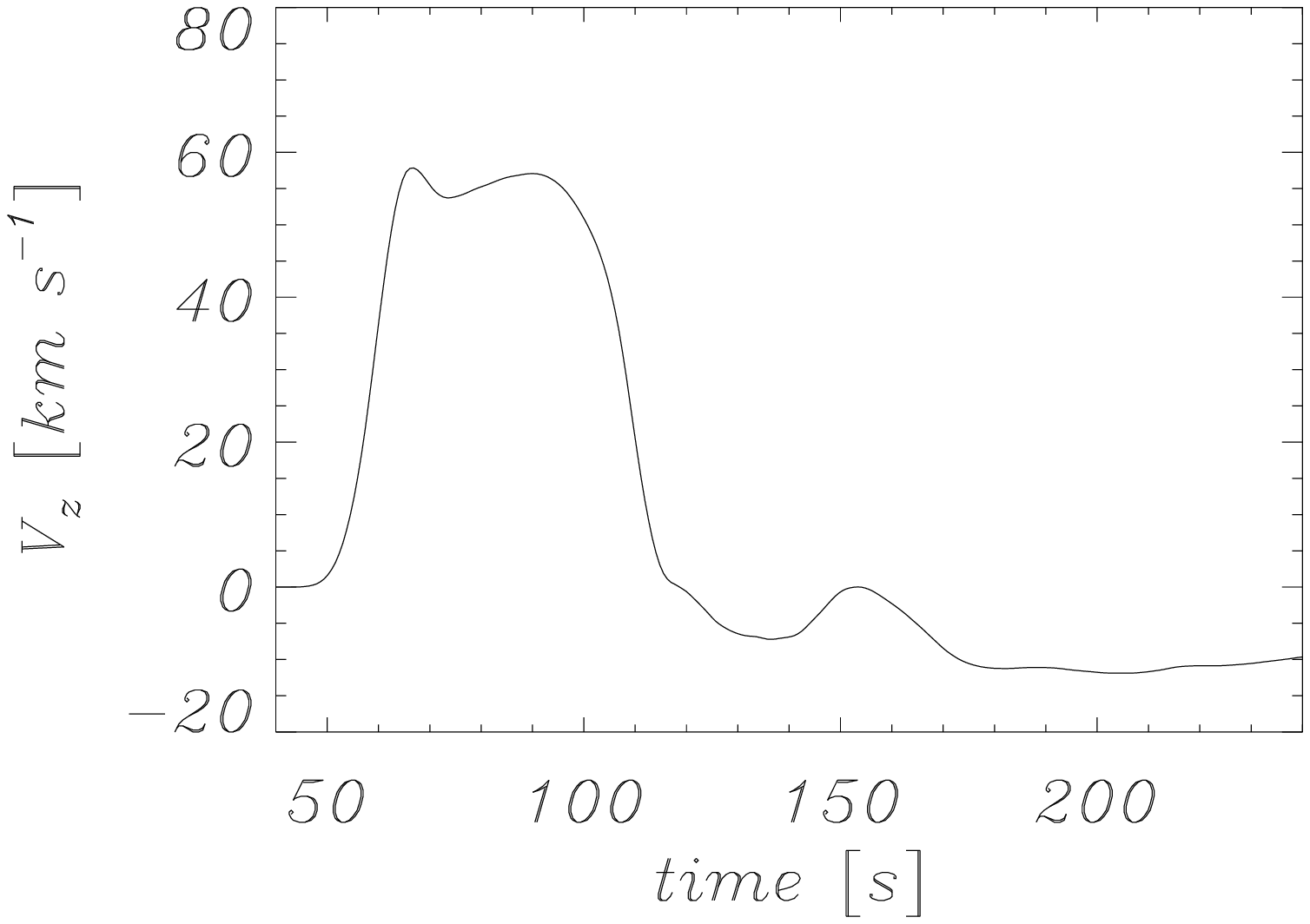}
}
\caption{\small
Results of the numerical
simulation of pulse-driven linear and non-linear Alfv\'en waves in the coronal funnels: Time-signatures of $\rho$, $V_{\rm y}$ and $V_{\rm z}$ for  
$A_{\rm v}=5$ km s$^{-1}$ (left column),
$A_{\rm v}=10$ km s$^{-1}$ (middle column)
and $A_{\rm v}=40$ km s$^{-1}$ (right column) 
collected at $(x=0, y=51)$ Mm.
} 
\label{fig:vz}
\end{center}
\end{figure*}

We study the case of an initial localized 
pulse that is launched in {$V_{\rm z}(x,y)$} at $t=0$~s perpendicular to the isosurface of 
the approximately open (weakly curved) and expanding magnetic field configuration.  
This pulse is described by Eq.~(\ref{eq:init_per}).
Examples of spatial profiles of $V_{\rm z}$ and the corresponding time-signatures 
obtained in our numerical simulation of Alfv\'en waves with different amplitudes are shown in Figs.~\ref{fig:6panels} 
and \ref{fig:vz}, respectively.  

%

First, we consider the pulse amplitude {$A_{\rm v}=5$ km~s$^{-1}$} for the simulation 
of the pulse-driven linear Alfv\'en waves in the coronal funnel.  Perturbations in 
$V_{\rm z}$ propagate along magnetic field lines, what is displayed on 
spatial profiles of $V_{\rm z}$ {at $t=22$~s} and $58$~s (Fig.~\ref{fig:6panels}).
A part of the whole simulation region is illustrated in the figure. 
The initial pulse launched at $y_{\rm 0}=1.75$~Mm decouples into two counter-propagating waves.
The amplitude of downwardly propagating waves falls off in time 
and these waves subside.
On the other hand, upwardly propagating Alfv\'en waves
experience an acceleration at the transition region as $c_A(y)$ increases up to 
$1$ Mm s$^{-1}$ there {(Fig.~\ref{fig:Tealf}, right)
and their profile become elongated along the vertical direction. 
Above the transition region, 
Alfv\'en waves propagate with a constant amplitude and velocity.
A part of the wave signal reflected at the transition region
is well seen at $y\sim2$~Mm at $t=58$~s (Fig.~\ref{fig:6panels}, right panel).


%

Panels in the left column of Fig.~\ref{fig:vz} illustrate typical features of the linear Alfv\'en wave.
Here, the profiles of the mass density and $y$-component of velocity
exhibit only small variations in time.
The growth of $V_{\rm y}$ is by two orders of magnitude smaller than transverse velocity
and it is directly associated with the Alfv\'en waves,
which reach the detection point $(x=0\, {\rm Mm}, y=51\, {\rm Mm})$
at $t\sim55$~s.
The mass density profiles exhibit a negligible decline during Alfv\'en wave passing.
It is significant that the Alfv\'en waves while pass through the solar atmosphere experience a compact 
shape 
(Fig.~\ref{fig:vz}, bottom panels).

The time-signature of 
$V_{\rm z}(x=0\, {\rm Mm}, y=51\, {\rm Mm})$  for $A_{\rm v}=5$ km~s$^{-1}$, 
illustrated in Fig.~\ref{fig:vz} ({bottom-left panel}), found to be similar to that 
obtained by Murawski \&~Musielak [29] [cf, their Fig.~6].
In the middle column of Fig.~\ref{fig:vz}, the time-signatures of Alfv\'en waves,
which are launched by the initial pulse with 
$A_{\rm v}$ = 10 km s$^{-1}$ are presented. 
Note that the time-signature of Alfv\'en waves for $A_{\rm v}=10$ km~s$^{-1}$ 
has a similar form as in the linear case.
%
However, the wave intensifies and enhances the plasma flows also along $y$-direction (Fig.~\ref{fig:vz}, middle-center panel)
two times 
in comparison to pulse with $A_{\rm v}=5$ km~s$^{-1}$.
We also simulate the Alfv\'en wave using the large transversal pulse
$A_{\rm v}=40$
km~s$^{-1}$ that generates the large-amplitude, non-linear Alfv\'en waves [22] 
to make comparison with Alfv\'en waves driven by two other pulses.  The time-signatures 
of V$_{z}$ {for $A_{\rm v}$ = 40 km s$^{-1}$  collected at ($x$ = 0, $y$ = 51) Mm are shown 
in the right column of Fig.~\ref{fig:vz}. 
This wave experiences non-linear effects,
as a result of which the vertical flows (Fig.~\ref{fig:vz}, middle-right panel) 
are driven by a ponderomotive force.

%

%
The time-signature of $V_{\rm z}$ exhibits a double peak shape.
This deformation 
arises as a result of partial reflection of the Alfv\'en waves
from the transition region.
The reflected signal reaches a dense photospheric plasma and reflects back towards the solar corona.
For a larger initial pulse amplitude this effect is more important.
Note, that even for Alfv\'en waves triggered by the pulse amplitude $A_{\rm v}$ = 10 km s$^{-1}$
the Gaussian shape of the wave signal is deformed
on its right side (Fig.~\ref{fig:vz}, bottom-middle panel).
In general, time-signatures of large-amplitude Alfv\'en waves exhibit similar behaviour
as of the linear Alfv\'en wave with $A_{\rm v}=5$ km~s$^{-1}$.
We observe effects of the non-linear Alfv\'en wave decoupling from magnetoacoustic waves
in $V_{\rm y}$
which magnitude grows with $A_{\rm v}$.
%
The spatial profile of $V_{\rm y}$ for $A_{\rm v}=40$ km~s$^{-1}$ (Fig.~\ref{fig:vz}, middle-right panel)
is like for $A_{\rm v}=5$ km~s$^{-1}$ or $A_{\rm v}=10$ km~s$^{-1}$,
although in the case (c) it is forty times bigger than in the linear case, (a)
and suffer deformation coming from a partial reflection of the Alfv\'en wave at the transition region.
Note that the mass density variations for Alfv\'en waves triggered by the initial
pulses with $A_{\rm v}=5$ km~s$^{-1}$, $A_{\rm v}=10$ km~s$^{-1}$ and $A_{\rm v}=40$ km~s$^{-1}$
(cf. Fig.~\ref{fig:vz}, top panels)
are insignificantly small, in particular for case (a) and (b),
which are two times smaller in comparison to the case (c).
%

%
}

\section{Discussion and Conclusions}
\label{sec:concl}
In our pulse driven Alfv\'en wave model, the pulse of a small and large amplitudes 
are launched above the solar photosphere in the weakly expanding coronal funnels.
The pulse splits into the upward and backward propagating wavetrains in the overlying 
solar atmosphere.  The upward moving pulse train results in the instantaneous displacement 
of the field lines in perpendicular plane away and towards the line-of-sight.  This 
effect generates the transverse velocity.  The pulse driven linear ($A_{\rm v}=5$ 
km~s$^{-1}$) and non-linear ($A_{\rm v}=10$ km~s$^{-1}$ and {$A_{\rm v}=40$ km~s$^{-1}$ }) 
Alfv\'en waves exhibit different contributions in the transverse velocity V$_{z}$ 
component at a particular height, and therefore they can play different roles in 
the local energy budget of the solar atmosphere.
A typical granular motions can freely trigger the linear Alfv\'en wave
with pulse amplitude $A_{\rm v}=5$ km~s$^{-1}$.
Whereas a generation of non-linear Alfv\'en wave is possible
in processes of microscopic magnetic reconnection like micro- or nano-flares,
where released kinetic energy disturb plasma in velocity generating MHD waves [38].
%

%

In theory, 
%
%
the maximum energy flux density carried out by such Alfv\'en waves under Wentzel-Kramers-Brillouin (WKB)
approximation is [39] 
\begin{equation}
F\approx
\rho \times V_z^2 \times C_{A}\, ,
\label{eq:amp}
\end{equation}
where $\rho$ is the density at the particular height, while the C$_{A}$ is the local 
phase speed of the  Alfv\'en waves. It should be noted that we do not calculate 
the energy flux distribution per degree of freedom, as we generalize the magnetic field
configuration where such modes are excited.
For the linear Alfv\'en waves ($A_{\rm v}=5$ 
km~s$^{-1}$) in the coronal funnel, and by considering
{
$\rho_{e}$=0.323$\times$10$^{-15}$ 
g cm$^{-3}$, V$_{\rm z}$=10 km s$^{-1}$ (cf., Fig.~\ref{fig:vz}, left column) as well as C$_{A}$=10${}^3$ 
km s$^{-1}$ 
}
at $y$=51~Mm (1.07 Ro) in the inner part of the coronal funnel, we 
can estimate the maximum energy flux carried out by such waves as
{
F$_{max}$$\approx$3.0$\times$10$^{4}$ ergs cm$^{-2}$ s$^{-1}$.
}

If the coronal funnel of the similar physical and magnetic field configuration 
exists in the polar coronal holes, then the energy carried out by such impulsively 
excited Alfv\'en waves will almost be sufficient to fulfill the energy requirement 
of the inner corona [39]. 
If the coronal funnel will be existing in the 
quiet-Sun or in form of active region loop arches, then such waves may only partially 
fulfill the energy requirements.  For the non-linear Alfv\'en waves ($A_{\rm v}=10$ 
km~s$^{-1}$) in the coronal funnels, and by considering 
{
$\rho_{e}$= 0.321$\times 
$10$^{-15}$ g cm$^{-3}$, V$_{\rm z}$=20 km s$^{-1}$ (cf., the middle column in Fig.~\ref{fig:vz}) 
as well as C$_{A}$=10${}^3$ km s$^{-1}$
}
at $y$=51 Mm (1.07 Ro) in its inner part, 
we can estimate the maximum energy flux carried out by such waves as 
{
F$_{max}$$\approx$1.3$\times$10$^{5}$ ergs cm$^{-2}$ s$^{-1}$.
}
This wave energy 
flux will be sufficient to fulfill the energy requirement of the inner corona [40]. 
However, for the coronal funnel that exists in form of active 
region loop arches, the computed wave energy flux only partially fulfills the 
required energy budget of the corona.

Now, for the non-linear Alfv\'en waves ($A_{\rm v}=40$ km~s$^{-1}$) in the coronal 
funnels, and by considering 
{
$\rho_{e}$= 0.325$\times$10$^{-15}$ g cm$^{-3}$, 
V$_{\rm z}$=60 km s$^{-1}$ (cf., Fig.~\ref{fig:vz}, right column) as well as C$_{A}$=10${}^3$ km 
s$^{-1}$
}
at a height of 51 Mm (1.07 Ro) in its inner part, we can estimate the 
maximum energy flux carried out by such waves as 
{
F$_{max}$$\approx$1.2$\times$10$^{6}$ ergs cm$^{-2}$ s$^{-1}$, 
}
which can again be sufficient to fulfill the energy requirement 
of the inner corona [40]. 
Chmielewski et al. [22] 
studied in details the role of such
non-linear Alfv\'en waves in the polar coronal holes and their role in the observed 
spectral line broadening. 

{
It should be noted that our calculation of the WKB part of the energy
are performed only to make comparison with the wave energy computed
for the linear waves.  However, applications of the WKB method are 
limited because of abrupt variations of the plasma parameters
in the stratified solar atmosphere whose spatial scales can be at a 
comparable magnitude to a typical wavelength.
%
Nevertheless, we apply the WKB method in the linear regime, i.e. we ignore the higher order terms,
which are related to higher-order mass density variations.
As a result, we even underestimate the energy flux evaluations
and the non-linear high-amplitude Alfv\'en waves are sufficient to fulfill the
localized coronal energy losses.
}

In conclusion, we showed that the linear pulse-driven Alfv\'en waves can sufficiently 
power the inner corona above the polar coronal holes ({for} $A_{\rm v}=5$ km~s$^{-1}$), 
while the non-linear waves {(for $A_{\rm v}=10$ km~s$^{-1}$ } carry enough energy to fulfill 
the energy requirement in solar coronal holes as well as quiet-Sun energy losses.  
Our results clearly demonstrated that non-linear Alfv\'en waves ($A_{\rm v}=40$ 
km~s$^{-1}$) can power the inner corona in the expanding coronal funnels existing 
anywhere, in the coronal holes, quiet-Sun, as well as in form of active region 
fan-like loop arches.  In general Alfv\'en wave dissipation occurs either in the 
distant part of the corona [41], 
or by some unique processes (e.g., 
phase-mixing, resonant absorption) in the solar atmosphere [42,43]. 
Therefore, the lower part of the solar atmosphere above the polar corona may be 
the ideal place for the undamped growth of linear and non-linear Alfv\'en waves, 
and sufficiently {large amplitude Alfv\'en} waves may supply (and damp) their energy in the 
outer part of the corona and can heat the solar wind ions [44]. 

{In the 
corona, 
}
the linear (e.g., $A_{\rm v}=5$ km~s$^{-1}$) and 
non-linear {Alfv\'en} waves (e.g., $A_{\rm v}=40$ km~s$^{-1}$) can carry sufficient 
{amount of} energy, and provide momentum to accelerate the solar wind plasma 
along the expanding magnetic field lines of the coronal funnels (e.g., 
coronal hole, and quiet-sun flux-tubes, respectively) [13]. 
Despite of polar 
coronal holes and quiet-Sun open {magnetic} field structures, the solar wind outflows 
and large-scale flows are also observed in the open magnetic arches near the 
boundary of the active regions [45]. 
Therefore, the large-amplitude, 
non-linear, pulse-driven Alfv\'en waves can provide sufficient energy to such 
kind of flows above the active regions in the expanding coronal funnels.

\medskip
\noindent {\bf Acknowledgments}
This work has been supported by NSF under the grant AGS 
1246074 (K.M. \& Z.E.M.) and by the Alexander von Humboldt Foundation
(Z.E.M.). 
The software used in this work was in part developed by the
DOE-supported ASCI/Alliance Center for Astrophysical Thermonuclear Flashes at the 
University of Chicago.
A.K.S. thanks Shobhna Srivastava for patient encouragement.
AKS acknowledges the financial support
from DST-RFBR-P117 project.
\medskip


\end{document}